\documentclass[a4paper,11pt]{article}
\pdfoutput=1 

\usepackage{jheppub} 

\usepackage[T1]{fontenc} 

\usepackage{comment}

\title{\boldmath Late time dynamics in SUSY saddle-dominated scrambling through higher-point OTOC}

\author[a,c]{Rathindra Nath Das,}
\author[b,d]{Sourav Dutta,}
\author[c]{Archana Maji}

\affiliation[a]{Institute for Theoretical Physics and Astrophysics and Würzburg-Dresden Cluster of Excellence
ct.qmat, Julius-Maximilians-Universität Würzburg,\\ 
Am Hubland, 97074 Würzburg, Germany}
\affiliation[b]{Department of Physics, Indian Institute of Technology Madras, \\
Chennai 600036, India }
\affiliation[c]{Department of Physics, Indian Institute of Technology Bombay,\\ Mumbai 400076, India}
\affiliation[d]{Center for Quantum Information, Communication and Computing
\\ IIT Madras, India}
\emailAdd{das.rathindranath@uni-wuerzburg.de}
\emailAdd{sourav@physics.iitm.ac.in}
\emailAdd{archana\_phy@iitb.ac.in}

\abstract{
In this article, we study the scrambling dynamics in supersymmetric quantum mechanical systems.  The eigenstate representation of such supersymmetric systems allows us to present an explicit form of the $2N$-point out-of-time-order correlator (OTOC) using two equivalent formalisms viz. ``Tensor Product formalism'' and ``Partner Hamiltonian formalism''. We analytically compute the $2N$-point OTOC for the supersymmetric 1D harmonic oscillator and find that the result is in exact agreement with that of the OTOC of the 1D bosonic harmonic oscillator system. The higher-point OTOC is a more sensitive measure of scrambling than the usual 4-point OTOC. To demonstrate this feature, we consider a supersymmetric sextic 1D oscillator for which the bosonic partner system has an unstable saddle in the phase space, which is absent in the fermionic counterpart. For such a system we show that the bosonic, the fermionic as well as the supersymmetric OTOC exhibit similar dynamics due to supersymmetry constraints. Finally, we illustrate the late-time oscillatory behaviour of \textit{higher-point OTOC} for saddle-dominated scrambling and anharmonic oscillator systems and propose \textit{it} to be a probe of late-time dynamics in non-chaotic systems that exhibit fast early-time scrambling. }

\begin{document} 
\maketitle
\flushbottom

\section{Introduction}
In recent times, the out-of-time-order correlator (OTOC) has been extensively used as a probe of operator scrambling in various branches of physics like condensed matter \cite{PhysRevB.107.L020202, Nakamura_2019, Roy_2021, Fan2017,qmbs1}, holographic theories \cite{Geng_2021,holo2, li2024outoftimeorder, PhysRevResearch.2.043399, Bhattacharyya2022}, Random matrix theory\cite{rm1,rm2,rm3}, quantum computation\cite{qc1,qc2,qc3,qc4} etc. 
OTOC acts as an indicator of the chaotic dynamics present in a system. 
The exponential growth of OTOC in the initial time indicates the existence of a positive Lyapunov exponent, which is a quantitative measure of chaos. 
A bound on the growth rate of quantum OTOC has been proved by  Maldacena, Shenker and Stanford (MSS) in the late time behaviour of the system \cite{MSS}. The aforementioned bound was initially motivated by the study of quantum information in black hole physics\cite{L5,L6,L7,L8,L10,L11}; Sachdev-Ye-Kitaev (SYK) model\cite{L12,L13,L14,L15} has also been found to satisfy it\cite{SYK}. This bound is a noteworthy distinction between classical and quantum chaos. In the quantum realm, the Lyapunov coefficient is confined by the MSS bound, whereas in classical scenarios, it can escalate to arbitrarily high values. 
A recent study established a more general infinite set of constraints on OTOC, with the MSS bound assuming prominence as the leading term \cite{Subleading}. In the Heisenberg picture, the 4-point OTOC for the operators $x$ and $p$, defined as $ C(t_1,t_2)=-\langle [x(t_1),p(t_2)]^2\rangle$ is a probe commonly used to analyse the chaotic nature of a system. The out-of-time-ordered nature of the correlators detects the effect of the measurement of one operator on the later measurement of the other. This can be seen explicitly by expanding the 4-point OTOC, 
\begin{equation}
 \begin{aligned}
    C(t_1,t_2)&=\langle p(t_2)x(t_1)x(t_1)p(t_2)\rangle+\langle x(t_1)p(t_2)p(t_2)x(t_1)\rangle-2\text{Re}[\langle x(t_1)p(t_2)x(t_1)p(t_2)\rangle]
\end{aligned}   
\end{equation}
The first two terms on the right-hand side of the above equation can be time ordered, and after the dissipation time scale of the system, they factorise as $\langle x(t_1)x(t_1)p(t_2)p(t_2)\rangle\sim \langle x(t_1)x(t_1)\rangle\langle p(t_2)p(t_2)\rangle$ and then they remain of the same order at a later time due to the isometry of time translation \cite{MSS}. On the other hand, the third term changes drastically as it does not get factorized in equal time correlators and in the presence of chaos, $x(t)$ perturbs the measurements more and more in a way that this term decays very rapidly. So, the rapid growth of OTOC is due to the rapid decay of the third term, which causes the existence of the positive Lyapunov coefficient. Considering this point and some mathematical advantages, Maldacena, Shenker, and Stanford \cite{MSS} introduced the regularized thermal OTOC as,
\begin{equation}
    F(t_1)= -\text{tr}[\sigma p(t=t_2)\sigma x(t=t_1)\sigma p(t=t_2)\sigma x(t=t_1)]
\end{equation}
with $\sigma^4=\frac{e^{-\beta H}}{\text{tr}(e^{-\beta H})}$ for some general Hermitian local operators $x$ and $p$. This definition of 4-point OTOC is very useful as it is analytic in a strip of width $\beta/2$ in the complex time plane, and in quantum field theory, it is free from coincident point singularities.

 In a semi-classical picture, we can replace the commutator $[x(t_1),p(t_2)]$ by Poisson bracket $i\hbar \{x(t_1),p(t_2)\}$, which is equivalent to $i\hbar\frac{\delta x(t_1)}{\delta x(t_2)}$. 
For a chaotic system with positive Lyapunov exponent $\lambda$, we can write $\frac{\delta x(t_1)}{\delta x(t_2)} \approx e^{\lambda |(t_1-t_2)|} $. It follows that the $2N$-point OTOC, which is the thermal average of $[x(t_1),p(t_2)]^{N}$, becomes equal to $e^{N\lambda t}$. So even if the value of the Lyapunov exponent is small, its effect gets magnified by a factor of $N$ for $2{N}$-point OTOC. The generalised $2N$-point OTOC can be regarded as a more sensitive measure of randomness present in the system.

However, we show here that solely examining the 4-point OTOC is insufficient to determine the chaotic properties of certain quantum systems. The 1D double well potential is one such example. Such a potential has an unstable saddle in the phase space, resulting in an early-time exponential growth of 4-point OTOC \cite{hasi}. Furthermore, it was shown that the quantum mechanical anharmonic oscillator having no unstable saddle can still give rise to fast scrambling and rapid growth of 4-point OTOC at early times \cite{Romatschke2021}.
In this work, we show that higher-point OTOCs are required to accurately capture a system's chaotic nature as they capture more fine-tuned correlations. We propose that the higher-point OTOC is a better probe for studying the late-time dynamics of the system, as after the early growth of OTOC, the higher-point correlations play a pivotal role in determining the late-time dynamics of the system. To illustrate this point, we examine a supersymmetric (SUSY) 1D sextic oscillator. In this system, the bosonic counterpart contains an unstable saddle point in the phase space, while the fermionic counterpart does not.

Supersymmetric quantum mechanics \cite{Cooper_1995, PhysRevD.106.046001} provides a way to go beyond the standard model physics by introducing a new symmetry transformation between the bosons and fermions. This symmetry relates particles with different spin properties, namely, fermions (spin-$1/2$ particles) and bosons (integer spin particles). In SUSY QM, for every known particle with an integer spin, there is a corresponding partner particle with a half-integer spin, and vice versa. The core idea of SUSY QM is to introduce a supersymmetric partner Hamiltonian that shares certain properties with the original Hamiltonian. Recent work has been done on the study of generalised OTOC in SUSY systems using Partner Hamiltonian formalism (PHF) in the eigenstate representation \cite{stn}. 

In this work, we choose to work in the eigenstate representation to extend the calculation of the OTOC for SUSY systems in Tensor Product formalism (TPF) as well as revise the work in PHF presented in the last work \cite{stn}. Our results for generalised OTOC in PHF differ from the results of the previous work \cite{stn}. Further, we present a way to calculate the generalised $2N$-point OTOC by following both formalisms. As a first example, we calculate the thermal $2N$-point OTOC for a 1D SUSY harmonic oscillator. The motivation behind choosing the 1D SUSY harmonic oscillator is that it is one of the few cases for which we can calculate the explicit analytical form of $2N$-point OTOC using both formalisms. From the results obtained for this system, we also verify that the results mentioned in the article \cite{stn} are inconsistent. This can be traced back to a wrong choice of SUSY identity operator in \cite{stn}.

Next, we consider the SUSY sextic oscillator system in the PHF. The bosonic Hamiltonian has an unstable saddle at $x=0$, whereas the fermionic oscillator has no unstable saddle but a higher anharmonic sextic potential. We study this setup to compare the growth of OTOC for saddle-dominated scrambling \cite{hasi} with the anharmonic potential effect \cite{Romatschke2021} in a SUSY system.  We find that the late-time behaviour of the higher-point OTOC of this saddle-dominated and anharmonic potential-driven system becomes oscillatory in nature. As we find, in the generalised definition of OTOC for the $2N$-point case, there will be many different terms other than of the form $\langle p(t_2)x(t_2)p(t_2)x(t_1)...p(t_2)x(t_1)\rangle$ that can not be factorized into equal time components and will contribute to the exponential growth of OTOC. We find that the OTOC growth in the saddle-dominated scrambling and the higher anharmonic potentials have the exact same behaviour due to the similar spectral properties governed by the constraints of SUSY quantum mechanics. Moreover, we find that the higher-point OTOC has a significant decay region followed by a highly oscillatory saturation region after the peak OTOC value following the initial growth.
\\
The rest of this paper is organized as follows. In section \ref{sec: TPF & PHF}, we briefly review both TPF and PHF in SUSY quantum mechanics and set the notations for the rest of the work. In section \ref{OTOC:TPF_PHF}, we define the generalised $2N$-point out-of-time-order correlator for supersymmetric systems using both TPF and PHF. We also consider an analytical example of the SUSY harmonic oscillator to highlight the corrections in the previous work \cite{stn}. Finally, in section \ref{sec: Saddle}, we study the higher-point OTOC for the SUSY sextic potential in PHF and propose the higher-point OTOC as a better probe of late-time dynamics.

\section{Preliminaries: supersymmetric quantum mechanics}
\label{sec: TPF & PHF}

In this section, we present a brief review of supersymmetric quantum mechanics\cite{susyqm}, explaining the notion of the tensor product formalism (TPF) and the partner Hamiltonian formalism (PHF).
The idea of a more generalized symmetry between bosons and fermions was introduced to address certain problems that the standard model of particle physics cannot explain.
This new symmetry allows for the transformation between a bosonic and a fermionic state. The tensor product and partner Hamiltonian formalisms are two equivalent ways to study systems with such symmetry.
\cite{Bagchi,Khare,Ramadevi,Notation}. 

\subsection{Tensor Product Formalism}
\label{rev:TPF_PHF}

In the TPF, initially, we define a tensor product Hilbert space of the bosonic and the fermionic systems. Let us denote the Hilbert space of the bosonic system by $\mathcal{H}_{B}$ and that of the fermionic system by $\mathcal{H}_{F}$. Then the SUSY Hilbert space would be the tensor product of these two Hilbert spaces viz., $\mathcal{H}_{S}=\mathcal{H}_{B}\otimes\mathcal{H}_{F}$.
Any supersymmetric quantum operator in the Hilbert space is defined as $\mathcal{O_S}\equiv\mathcal{O_B}\otimes\mathcal{O_F}:\mathcal{H_B}\otimes\mathcal{H_F}\rightarrow\mathcal{H_B}\otimes\mathcal{H_F}$,  where $\mathcal{O_B},~\mathcal{O_F}$ acts on bosonic and fermionic systems respectively. Bosonic annihilation and creation operators are denoted as $a$ and $a^{\dagger}$. The quantization of the bosonic system requires commutation relation to be imposed between $a$ and $a^{\dagger}$ as 
\begin{align}
    [a, a^{\dagger}] =1. 
\end{align}
As for the fermionic system, the annihilation and creation operators are denoted as $c$ and $c^{\dagger}$. Due to the Pauli's exclusion principle imposed on fermions, $c$ and $c^{\dagger}$ must satisfy the following anticommutation relation
\begin{align}
    \{c, c^{\dagger}\} =1. 
\end{align}
The number operators for bosonic and fermionic systems are $N_{B}=a^{\dagger}a$, $N_{F}=c^{\dagger}c$. These operators acting on the n-th number state give $N_{B}|n_{B}\rangle =n_{B}|n_{B}\rangle$, where, $n_{B}=0,1,2,...$ and, $N_{F}|n_{F}\rangle =n_{F}|n_{F}\rangle$, where, $n_{F}=0,1$.
We can represent any supersymmetric state as $|n_B\rangle\otimes |n_F\rangle$ in the tensor product formalism. 
We are all set to define the supercharge operator on supersymmetric Hilbert space. The supercharge operator $Q= a \otimes c^{\dagger}$ converts a boson into a fermion. On the other hand, the conjugate supercharge operator  $Q^{\dagger}=a^{\dagger}\otimes c$ converts a fermion into a boson.
\begin{equation}
\begin{aligned}
    Q|n_{S}\rangle&=(a\otimes c^{\dagger})|n_{B}\rangle \otimes |n_{F}\rangle=|(n-1)_{B}\rangle \otimes |(n+1)_{F}\rangle\\
    Q^{\dagger}|n_{S}\rangle&=(a^{\dagger}\otimes c)|n_{B}\rangle \otimes |n_{F}\rangle=|(n+1)_{B}\rangle \otimes |(n-1)_{F}\rangle .
\end{aligned}
\end{equation}
Since $Q$ and $Q^{\dagger}$ generate the states of the supersymmetric system, these are called generators.
In the following section, we elaborate on the formulation of OTOC using PHF.
\subsection{Partner Hamiltonian Formalism}
The partner Hamiltonian formalism provides a powerful mathematical tool to investigate the symmetries and properties of SUSY quantum systems using two partner Hamiltonians. 
As the SUSY generates transformation between a bosonic and a fermionic state, one can define two separate but related Hamiltonians governing these two systems.
Referred to as the bosonic and fermionic Hamiltonians, the eigenstates of these two Hamiltonians are superpartners of each other and share similarities in all quantum numbers except for the spin quantum number.
The most general time-independent Schrodinger equation of the first system can be written as 
 \begin{align}
     H_1\psi_{1}^{(n)}=- \frac{\hbar^2}{2m}\frac{d^2}{dx^2}\psi_{1}^{(n)}+V_1(x)\psi_{1}^{(n)}=E_{1}^{(n)}\psi_{1}^{(n)}.
 \end{align}
 Here $\psi_{1}^{(n)}$ and $E_{1}^{(n)}$ are the n-th eigenstate and eigenvalue of the Hamiltonian $H_1$. The potential $V_1(x)$ can be expressed in terms of the ground state and its corresponding energy,  
 \begin{align}
     V_1(x)=\frac{\hbar^2}{2m}\frac{1}{\psi_{1}^{(0)}}\frac{d^2\psi_{1}^{(0)}}{dx^2}+E_{1}^{(0)}.
 \end{align}
We define the bosonic Hamiltonian by subtracting the ground state energy from the first Hamiltonian to keep zero ground state energy for the bosonic system as $ H_B=H_1-E_1^{(0)}=- \frac{\hbar^2}{2m}\frac{d^2}{dx^2}+V_B(x)$,
where, $V_B = \frac{\hbar^2}{2m}\frac{1}{\psi_{1}^{(0)}}\frac{d^2\psi_{1}^{(0)}}{dx^2}$. 
In order to define the partner fermionic Hamiltonian, one must determine the fermionic partner potential. This necessitates actively defining a superpotential, denoted as $W(x)$, which adheres to the relationship 
\begin{align}
   & V_B = W^{2}(x)-\frac{\hbar}{\sqrt{2m}}W^{\prime}(x)\label{n8}\\
   & V_F = W^{2}(x)+\frac{\hbar}{\sqrt{2m}}W^{\prime}(x)\label{n9}
\end{align}
Using these equations, for any known $V_B$, one can determine $V_F$ as well as the fermionic Hamiltonian, $H_F = - \frac{\hbar^2}{2m}\frac{d^2}{dx^2}+V_F$.
The n-th eigenstates of $H_B$ and $H_F$ are $\psi_{B}^{(n)}$ and $\psi_{F}^{(n)}$ respectively with corresponding eigenvalues $E_{B}^{(n)}$ and $E_{F}^{(n)}$. 
However, the bosonic and fermionic energy eigenvalues follow the relation 
\begin{equation}
 E_{B}^{(n+1)}=E_{F}^{(n)}.  \label{susyE}   
\end{equation}

As a result, we can represent the eigenstate of the SUSY system $\psi_{S}^{(n)}$ as, 
\begin{align}
    \psi_{S}^{(n)}=\begin{cases}
    \frac{1}{\sqrt{2}}& \begin{bmatrix}
    \psi_{B}^{(n)}\\
    \psi_{F}^{(n-1)}
    \end{bmatrix}~~~ \text{for } n \neq 0 \\
    \\
    &\begin{bmatrix}
    \psi_{B}^{(0)}\\
    0
    \end{bmatrix} ~~~ \text{for } n = 0
    \end{cases} \label{state}
\end{align}
where $\psi_{S}^{(0)}$ is the ground state of the SUSY system. 
It is worth noticing that the ground state with zero energy is the only state without a partner fermionic state. 
The relations between these Partner Hamiltonian states and the Tensor Product states have been discussed in detail in \cite{rnda}.
\section{OTOC in supersymmetric systems}
\label{OTOC:TPF_PHF}
In this part, we define the higher-point OTOC in supersymmetric quantum mechanics using both of the formalisms discussed above. Then, we consider one example that can be treated using both TPF and PHF analytically, namely the SUSY 1D harmonic oscillator. Another motivation for choosing this example is to show the consistency of our results across two formalisms, underscoring the potential need for adjustments to the findings outlined in \cite{stn}.

\subsection{Formulation of OTOC in TPF}
 We define the generalised $2N$-point out-of-time-order correlator as the thermal average of the $N$-th power of two non-commutating operators at two different times.
 \begin{align}
 C_{T}(t_1,t_2)=(-i)^N\langle[x(t_1),p(t_2)]^{N}\rangle \label{ab0}
 \end{align}
 Here, we have chosen the position operator at time $t_1$ and the momentum operator at time $t_2$ as non-commuting operators.
 The presence of the term $(-i)^N$ seems to result in the thermal OTOC becoming a real quantity that is directly proportional to the exponential function $e^{N\lambda t}$.
 In this case, the value of $N$ can be any positive integer, and we can adjust it as a parameter to observe the amplified impact of quantum chaos. We will now derive the expression for the 4-point OTOC using the tensor product framework and generalize the expression to encompass the $2N$-point OTOC.
When represented in the eigenstate format, the OTOC takes the following structure:
\begin{align}
 C_{T}(t_1,t_2)=&\frac{1}{Z} \sum_{n} e^{-\beta E_{n}} c_{n}(t_1,t_2) \label{a24}
\end{align}
In Eq. \eqref{a24}, $c_{n}$ denotes the microcanonical OTOC, while $C_T$ is referred to as thermal OTOC. 
For the 4-point OTOC, the microcanonical OTOC has the following form
\begin{equation}
    c_{n}(t_1,t_2)=-\langle n_{B},n_{F}|[x(t_{1}),p(t_{2})]^{2}|n_{B},n_{F} \rangle .\label{ab1}
\end{equation} 
Here, $\beta = \frac{1}{k_BT}$ and $T$ is the temperature for the ensemble. $Z=\sum_ne^{-\beta E_{n}}$ is the thermal partition function of the system in the eigenstate representation.
Eq. \eqref{ab1} can be expanded as,
\begin{align}
c_{n}(t_1,t_2)=&(-i)^2\langle n_{B},n_{F}|[x(t_{1}),p(t_{2})]^{2}|n_{B},n_{F} \rangle = \sum_{k}d_{nk}(t_1,t_2)d_{kn}(t_1,t_2) \label{ab2}\\
=& \frac{1}{4}\sum_{k,m} \big|x_{nm}x_{mk}(e^{iE^{nm}t_1}e^{iE^{mk}t_2}E^{mk}-e^{iE^{nm}t_2}e^{iE^{mk}t_1}E^{nm})\big|^2\label{a40}
\end{align}
where,
\begin{align}
d_{nm}(t_1,t_2)&=-i\langle n_{B},n_{F}|[x(t_{1}),p(t_{2})]|m_{B},m_{F} \rangle   ,~ x_{nm}=\langle n_{B},n_{F}|x(t=0) |m_{B},m_{F} \rangle ,\nonumber\\
&~~~~~~~~~~~~~~~~~~~~~~~E^{nm}=E^n-E^m.\nonumber
\end{align}
 The expression \eqref{a40} can be substituted in \eqref{a24} to obtain the 4-point OTOC.
The extension from the $4$-point OTOC to the $2N$-point OTOC is a relatively straightforward process. 
We can express the microcanonical $2N$-point OTOC as,
\begin{align}
c_{n}(t_1,t_2)=(-i)^N\langle n_{B},n_{F}|[x(t_{1}),p(t_{2})]^{N}|k_{B},k_{F}\rangle=\sum_{k_1..,k_{(N-1)}}d_{nk_{1}}d_{k_1k_2}...d_{k_{N-1}n}. \label{a31}
\end{align}  
After some calculations, we obtain the following expression for $d_{nm}(t_1,t_2)$ 
\begin{align}
d_{nm}(t_1,t_2)= \frac{1}{2} \sum_{k} x_{nk}x_{km}(e^{iE^{nk}t_1}e^{iE^{km}t_2}E^{km}-e^{iE^{nk}t_2}e^{iE^{km}t_1}E^{nk}). \label{a32}
\end{align}
It's important to note that Eq. \eqref{a32} serves as a connection between Eq. \eqref{ab2} and Eq. \eqref{a31}.
We can calculate the value of $d_{nm}$ for any given values of $n$ and $m$ using Eq. \eqref{a32}. These values can then be used to compute the $2N$-point microcanonical OTOC, $c_n(t_1,t_2)$. Finally, by employing Eq. \eqref{a24}, we can obtain the $2N$-point thermal OTOC.

\subsection{Formulation of OTOC in PHF}
By employing the prescribed definition of the $2N$-point OTOC, outlined in Eq. \eqref{ab0} and \eqref{a24}, we can articulate the microcanonical OTOC within the framework of the eigenstate representation of the PHF as,
\begin{align}
    c_{n}(t_1,t_2)=(-i)^N\langle \psi_{S}^{(n)}|[x(t_1),p(t_2)]^{N}|\psi_{S}^{(n)}\rangle \label{c1}
\end{align}
We can express the SUSY identity $\mathcal{I}_S$ in terms of $\mathcal{I}_B$ and $\mathcal{I}_F$, where $\mathcal{I}_B$ stands for the bosonic identity and $\mathcal{I}_F$ stands for the fermionic identity.
\begin{align}
  \mathcal{I}_S &=\begin{bmatrix}
   \mathcal{I}_B & 0 \\
   0 & \mathcal{I}_F
   \end{bmatrix} 
   =\begin{bmatrix}
   \sum_{k}|\psi_{B}^{(k)}\rangle\langle\psi_{B}^{(k)}| & 0 \\
   0 & \sum_{k}|\psi_{F}^{(k)}\rangle\langle\psi_{F}^{(k)}|
   \end{bmatrix} .\label{c2}
\end{align}

By utilizing the SUSY identity presented in Eq. \eqref{c2}, we can expand the microcanonical OTOC $c_n$ as defined in Eq. \eqref{c1}. Through subsequent intermediate steps, we arrive at the final expression for the microcanonical OTOC within the PHF, which can be succinctly represented as,
\begin{align}
  c_n(t)=&
\begin{cases}
\frac{1}{2}\sum\limits_{k_1,k_2,..,k_{N-1}}(b_{nk_1}b_{k_1k_2}...b_{k_{N-1}n}+f_{(n-1)k_{1}}f_{k_{1}k_{2}}...f_{k_{N-1}(n-1)}),~\text{for}~n\neq0\\
\sum\limits_{k_1,k_2,..,k_{N-1}}(b_{0k_1}b_{k_1k_2}...b_{k_{N-1}0}),~\text{for}~n=0
\end{cases}  
\label{n17}
\end{align}

where,

{
\begin{equation}
\begin{aligned}
 b_{nm}&= \sum_{k}\frac{1}{2}x_{nk}x_{km}(e^{iE_{B}^{nk}t_1}e^{iE_{B}^{km}t_2}E_{B}^{km}-e^{iE_{B}^{nk}t_2}e^{iE_{B}^{km}t_1}E_{B}^{nk}),\\ f_{nm}&=\sum_{k}\frac{1}{2}x_{nk}x_{km}(e^{iE_{F}^{nk}t_1}e^{iE_{F}^{km}t_2}E_{F}^{km}-e^{iE_{F}^{nk}t_2}e^{iE_{F}^{km}t_1}E_{F}^{nk}),\\
 &~~\text{with}~~E_{B}^{nk} =E_{B}^{n}-E_{B}^{k}.\label{t18}
\end{aligned}
\end{equation}}
Upon substituting the determined values of $b_{mn}$ and $f_{nm}$ from Eq. \eqref{t18} into Eq. \eqref{n17}, we are able to derive the value of the $2N$-point microcanonical OTOC $c_n(t)$. 
Consequently, we can proceed to perform a thermal averaging process across all the microcanonical OTOCs, as indicated by Eq. \eqref{a24}, leading to the final expression for the thermal OTOC corresponding to a specific temperature.

\subsection{Analytical example: harmonic oscillator}
The harmonic oscillator potential holds a distinctive significance within SUSY quantum mechanics due to its property of being shape-invariant.
The energy levels of the harmonic oscillator exhibit equal spacing.
Moreover, Heisenberg-Weyl algebra, a fundamental mathematical framework, can be effectively realized using such potentials. 
We shall investigate the OTOC for the harmonic oscillator potential using both TPF and PHF.\\
In the harmonic oscillator system, the energy difference between two states is given by the expression $E^{nm}=(n-m)\hbar\omega$. Additionally, the matrix element $x_{nm}$ can be expressed as shown in Eq. \eqref{b1}.

{
\begin{equation}
\begin{aligned}
    &x_{nm}=\frac{1}{\sqrt{2}} (\sqrt{n_B+1}\delta_{n_B+1,m_B}\delta_{n_F,m_F}+ \sqrt{n_B}\delta_{n_B-1,m_B} \delta_{n_F,m_F}\\
    &~~~~~~~~~~~~~~~~+ \delta_{n_B,m_B}\delta_{n_F+1,m_F}+ \delta_{n_B,m_B}\delta_{n_F-1,m_F}). \label{b1}
\end{aligned}
\end{equation}}

By substituting Eq. \eqref{b1} into Eq. \eqref{a32} and performing a series of calculations, we find that $d_{nm}(t_1,t_2)= \cos(t_1-t_2)\delta_{nm}$. Subsequently, from Eq. \eqref{a31}, we deduce that the microcanonical $2N$-point OTOC $c_n(t_1,t_2)$, takes the form $c_{n}(t_1,t_2)=\cos^{N}(t_1-t_2)$. It is worth noting that $c_n(t_1,t_2)$ does not depend on the specific value of $n$ for the harmonic oscillator potential.
The thermal $2N$-point OTOC, $C_{T}(t_1,t_2)$, can be obtained from Eq. \eqref{a24} and is given by 
\begin{align}
 C_{T}(t_1,t_2)=\cos^{N}(t_1-t_2). \label{b2}
 \end{align}
This result demonstrates that the thermal OTOC oscillates around zero, indicating the absence of chaotic behaviour. This distinct oscillatory thermal OTOC is a consequence of the unique characteristics of the harmonic oscillator potential. 
For other potential systems, the behaviour of higher-point OTOC may vary compared to lower-point OTOC. This is because the higher-point OTOCs provide a more sensitive measure to the presence of chaos within the system, which we will explore in section \ref{sec: Saddle}.

We shall examine the $2N$-point thermal OTOC pertaining to the simple harmonic oscillator system within the partner Hamiltonian formalism.
The microcanonical OTOC for a SUSY system in PHF can be expressed by means of Eq.\eqref{n17}. 
The $b_{nm}$ of $2N$-point OTOC  for arbitrary $m$ and $n$ is given by 

\begin{align}
   b_{nm} &=-i  \sum_{k}(e^{iE_{B}^{nk}t_1}e^{iE_{B}^{km}t_2}x_{nk}p_{km}-e^{iE_{B}^{nk}t_2}e^{iE_{B}^{km}t_1}p_{nk}x_{km})\nonumber\\
   &=\cos(t_1-t_2)(\sqrt{(n+1)(m+1)}-\sqrt{nm})\delta_{n,m},
   \end{align}
   where we have replaced $p_{nm}$ with $\frac{i}{2}x_{nm}E_{nm}$.
 It is noteworthy that the wavefunction forms for both bosonic and fermionic states are identical. 
 Consequently, we can assert that $f_{nm}$ shares the same structure as $b_{nm}$. 
 Utilizing Eq. \eqref{n17}, we obtain the microcanonical OTOC for the SUSY harmonic oscillator utilizing the PHF, which manifests as $c_n(t_1,t_2)=\cos^{N}(t_1-t_2)$. 
 Moreover, the $2N$-point thermal OTOC within the PHF can be expressed as
\begin{align}
    C_T(t_1,t_2)=\cos^{N}(t_1-t_2). \label{n18}
\end{align}

 From Eq. \eqref{b2} and Eq. \eqref{n18}, we achieve an identical analytical expression for the $2N$-point OTOC using the TPF and PHF, respectively. 
 Notably, we observe that for $N=2$, we obtain $C_T(t_1,t_2)=\cos^{2}(t_1-t_2)$, coinciding with the thermal OTOC of the one-dimensional bosonic harmonic oscillator \cite{hasi}. 
 It is crucial to emphasize that this finding differs from the outcomes presented in Eq. 149 and Eq. 151 of \cite{stn}. Furthermore, the microcanonical OTOC stated in the same reference also deviates from our findings. 

\section{Saddle-dominated scrambling, anharmonic oscillators}
\label{sec: Saddle}

In this section, we consider the SUSY sextic oscillator with a special pair of supersymmetric partner potentials. The bosonic potential contains an unstable saddle in the phase space, while the fermionic partner potential does not. We demonstrate that for such a system, both the bosonic and fermionic parts show a similar initial growth of OTOC, which is also reflected in the OTOC growth of the whole SUSY system. This serves as a special case where the saddle-dominated scrambling has dynamics similar to that of an anharmonic sextic oscillator in terms of OTOC. We further study the higher-point OTOC for this system and show oscillatory dynamics in the late-time regime of the system, showcasing the non-chaotic nature of the system considered.

\begin{figure}[hbtp]
\begin{center}
\includegraphics[width=0.85\textwidth]{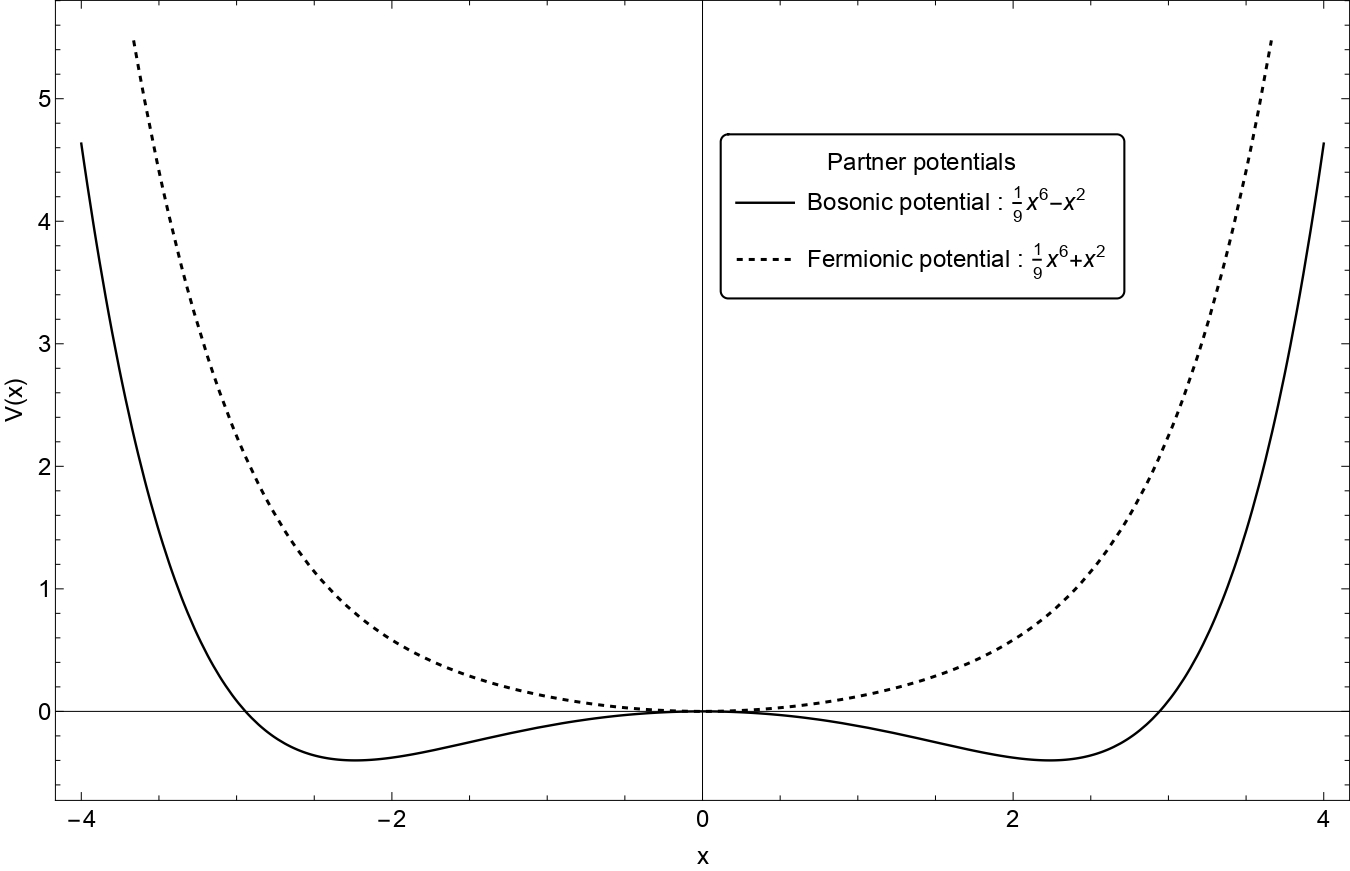}
\end{center}
\caption{Supersymmetric partner potentials for bosonic and fermionic systems. In contrast to the Fermionic case, there exists an unstable maxima in the Bosonic case.}
\label{worldlineofprth}
\end{figure}

\subsection{Generalised OTOC and late-time behaviour}

The fast scrambling nature of this system, which is justified by the rapid initial growth followed by a saturation phase of the four-point OTOC change drastically as we consider higher-point OTOC. The generalised OTOC reveals an oscillatory non-chaotic nature of saturation at late times and confirms that these systems are indeed integrable. The higher-point OTOC shows a significant decay region after the peak which is not very prominent in the four-point OTOC.  This observation highlights the necessity of considering not only the early-time growth but also the late-time behaviour when studying the OTOC, in order to gain a comprehensive understanding of the true nature of the system.

To see the late-time dynamics of the higher-point generalised OTOC, we take the supersymmetric system with the following partner potentials,
\begin{align}
\begin{split}
    V_B(x)&=16B^2x^6-12Bx^2\\
    V_F(x)&=16B^2x^6+12Bx^2
\end{split}
\end{align}
where $B$ is a free parameter. $V_B(x)$ and $V_F(x)$ are the bosonic and the fermionic partner potentials respectively. The bosonic partner potential has the ground state, $\psi(x)=Ae^{-Bx^4}$, with $A$ as the normalisation constant. For our purpose, we consider $B=1/12$. Notably, this is a very special pair of partner potentials as the bosonic part has an unstable saddle point at $x=0$, and the fermionic partner does not have any unstable saddle but has an anharmonic term as shown in Fig. \ref{worldlineofprth}. 

\begin{figure}[hbtp]
\begin{center}
\includegraphics[width=0.85\textwidth]{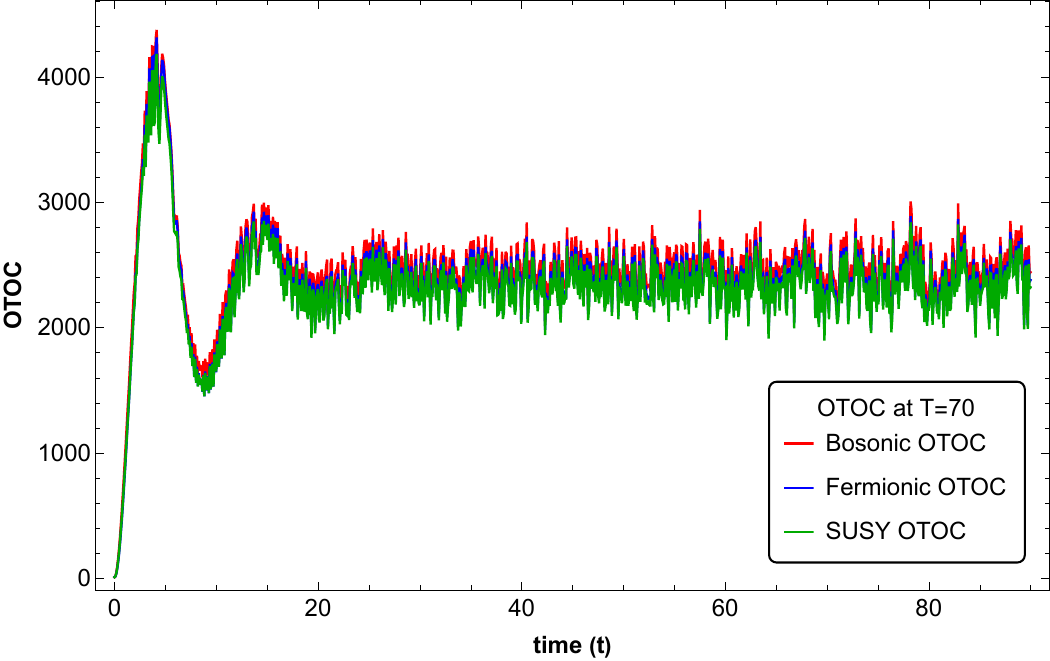}
\end{center}
\caption{Plots of N=2 OTOC at T=70 vs time for the Bosonic, Fermionic and supersymmetric systems. Despite having different potentials in all three cases, the OTOCs behave in a qualitatively similar fashion. }
\label{BFS_otoc}
\end{figure}
\begin{figure}[hbtp]
\begin{center}
\includegraphics[width=0.85\textwidth]{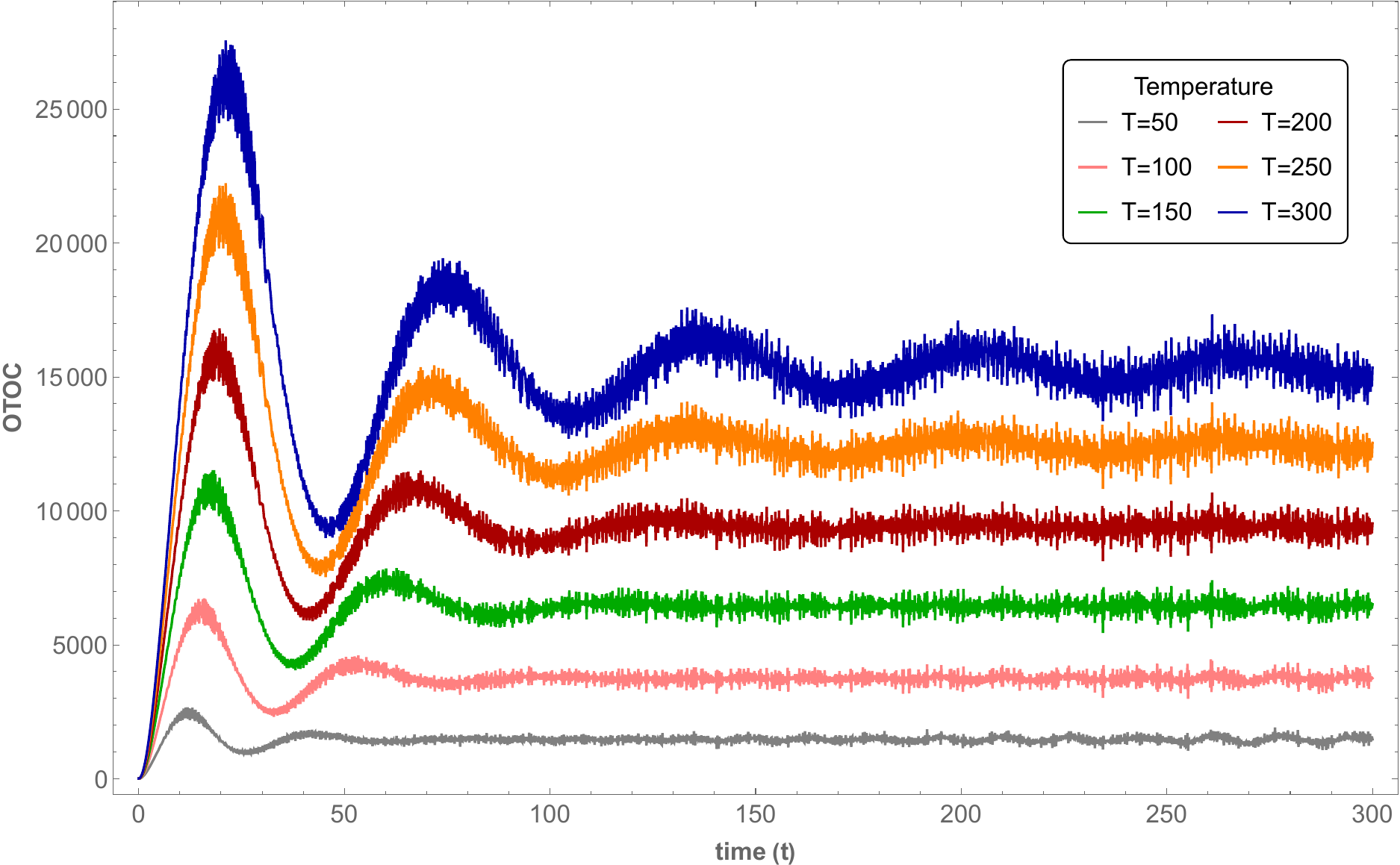}
\end{center}
\caption{Plots of N=2 OTOC vs. time for different temperatures. The peak, saturation value and growth rate of OTOC increase with temperature, showcasing faster scrambling behaviour. }
\label{otoc_diffT}
\end{figure}

\begin{figure}[hbtp]
\begin{center}
\includegraphics[width=0.85\textwidth]{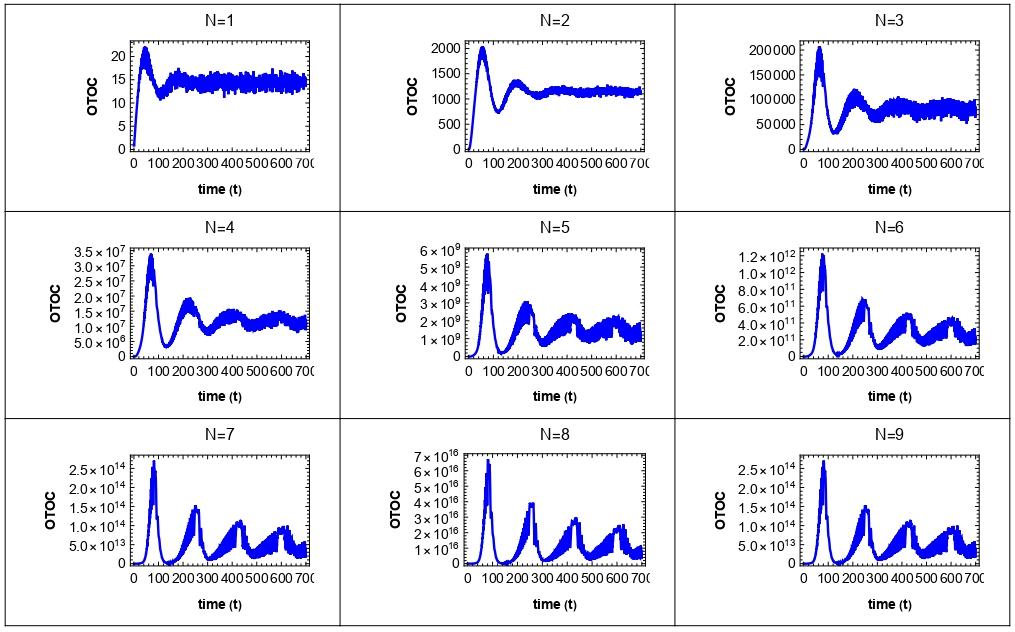}
\end{center}
\caption{Plots of $2N$-point OTOC at $T=200$ vs. time for different values of $N$ defined in \eqref{ab0}. The decay of OTOC after the initial growth and peak becomes more prominent for higher-point OTOC. The late-time oscillatory saturation dynamics highlight the non-chaotic nature of the system, which is undetectable from the conventional 4-point OTOC.}
\label{2NptOTOC}
\end{figure}

We find that the OTOC in the bosonic, fermionic and supersymmetric system behaves the same way in Fig. \ref{BFS_otoc}. This behaviour can be attributed to the spectra of the partner Hamiltonians, which follow \eqref{susyE}. This example serves as special case where the saddle-dominated scrambling can have the similar dynamics as the higher anharmonic potentials without a saddle. We further notice that after the initial growth and peak of the OTOC, a decay region is present before late-time saturation. We propose this as a signature of OTOC in systems without chaotic dynamics but saddle-dominated or anharmonicity-driven initial growth.

Next, we focus only on the SUSY OTOC as it has similar dynamics to those of bosonic and fermionic partner systems. In Fig. \ref{otoc_diffT}, we plot four-point OTOC for different temperatures. The peak and saturation value of OTOC increases for higher temperature values. The growth rate of OTOC is also proportional to the temperature.

Finally, we study the $2N$ OTOC for $N=1$ till $N=9$ in Fig. \ref{2NptOTOC}. Here $N=9$ corresponds to the 18-point OTOC or the 9th moment of OTOC. We find a striking behaviour here that for the higher-point functions, the late time saturation is highly oscillatory, and the dip after the peak becomes steeper. This can be understood as the late-time dynamics of this one-dimensional system being non-chaotic. After the initial growth of OTOC due to the unstable saddle, when the system reaches a steady condition, the OTOC shows an oscillatory behaviour. We propose this behaviour of higher-point OTOC as a generic behaviour for the non-chaotic systems where initial growth is due to saddle-dominated scrambling and anharmonic potential. In the 4-point OTOC, the main non-trivial contribution comes from $\langle x(t)p(0)x(t)p(0)\rangle$ with $t=|t_1-t_2|$, but in the higher-point OTOC other non-trivial contributions appear, e.g., $\langle x(t)p(0)x(t)p(0)x(t)p(0)x(t)p(0)\rangle$,  $\langle x(t)p(0)p(0)x(t)x(t)p(0)x(t)p(0)\rangle$, $\langle x(t)p(0)x(t)p(0)p(0)x(t)p(0)x(t)\rangle$, $\langle p(0)x(t)p(0)x(t)x(t)p(0)x(t)p(0)\rangle$ for the 8-point OTOC. These correlations contribute to the late-time decay and oscillations of OTOC.

\section{Conclusion}
In this article, we extend the study of quantum chaos for supersymmetric systems. The generalised $2N$-point OTOC for a SUSY system is formulated using both the Tensor Product and Partner Hamiltonian formalisms in the eigenstate representation.  This generalisation enables us to study more minute changes in the OTOC dynamics as it offers higher-order non-trivial correlations compared to the conventional 4-point OTOC. Important time scales like scrambling time, dissipation time, and Ehrenfest time can be defined as in the usual non-SUSY case. 

An explicit example of a supersymmetric 1D harmonic oscillator system has been considered to show that the $2N$-point thermal OTOC has a form of $\cos^N(t_1-t_2)$ for both the formalisms and fixes the inconsistencies of the previous work \cite{stn}.  With this extension of the definition of OTOC in SUSY quantum mechanics, we initiate a study of scrambling in saddle-dominated and anharmonic SUSY systems. We find that for the specific SUSY sextic potential that we consider, both the bosonic and fermionic sectors have a similar OTOC growth as the whole SUSY system due to the spectra of each sector being constrained by the supersymmetry. This is a special system where the saddle-dominated scrambling shows behaviour similar to that of a system with anharmonic potential. Furthermore, we find that a steep decay region is present in the 4-point OTOC in this system, which is more enhanced in the higher-point OTOC. We calculate till 18-point OTOC for this system, where we find that the late time oscillation becomes more prominent as we calculate the higher-point correlations. This further highlights the fact that this system is inherently non-chaotic, and the initial growth in the bosonic and fermionic partner systems is due to the presence of either an unstable saddle or the anharmonic term in the potential only. Further investigation of these higher-point OTOCs is needed to understand better the late-time dynamics of such systems. However, here we propose that the higher-point OTOC can serve as a tool to understand the scrambling dynamics of the system in the late time regime and possibly distinguish integrable systems.

Though we have used the conventional definition of OTOC to generalise for the $2N$-point case, following the MSS prescription, another way of generalising the definition of the $2N$-point can be as follows,
\begin{equation}
    F(t)= \text{tr}[\sigma p(t_2) \sigma x(t_1) \sigma p(t_2) \sigma x(t_1)... \sigma p(t_2)yx(t_1)]
\end{equation}
with $N$ number of $p(t_2)$ and $x(t_1)$ operators and $\sigma^{2N}=\frac{e^{-\beta H}}{\text{tr}(e^{-\beta H})}$.  How the results differ for the generalised $2N$-point OTOC for a chaotic system for these two definitions will be an interesting point to investigate in future.
This new generalisation is a useful tool for understanding the scrambling dynamics present in a quantum system.
\raggedbottom

\acknowledgments

We are grateful to Prof. Dr. Urjit A. Yajnik for his continuous support and helpful advice during the project. We would like to thank Dr. Johanna Erdmenger for her valuable suggestions. We would also like to thank Dr. Sayantan Choudhury for the training received during his earlier work with him, which has been, to an extent, helpful in completing this research.

\bibliographystyle{JHEP}
\bibliography{ref}

\end{document}